\documentclass[utf8]{FrontiersinHarvard} 

\usepackage{url,hyperref,microtype,subcaption}
\usepackage[onehalfspacing]{setspace}
\usepackage[T1]{fontenc}

\def\keyFont{\fontsize{8}{11}\helveticabold }
\def\firstAuthorLast{W\"unsch} 
\def\Authors{Richard W\"{u}nsch\,$^{1,*}$}
\def\Address{$^{1}$Astronomical institute of the Czech Academy of Sciences, Prague, Czech Republic
}
\def\corrAuthor{Richard W\"{u}nsch,\\
Astronomical institute of the Czech Academy of Sciences, Bo\v{c}n\'\i\ II 1401, CZ-14100 Prague, Czech Republic}

\def\corrEmail{richard.wunsch@asu.cas.cz}

\begin{document}
\onecolumn
\firstpage{1}

\title {Radiation transport methods in star formation simulations} 

\author[\firstAuthorLast ]{\Authors} 
\address{} 
\correspondance{} 

\extraAuth{}

\maketitle

\begin{abstract}

\section{}

Radiation transport plays a crucial role in star formation models, as certain questions within this field cannot be accurately addressed without taking it into account. Given the high complexity of the interstellar medium from which stars form, numerical simulations are frequently employed to model the star formation process. This study reviews recent methods for incorporating radiation transport into star formation simulations, discussing them in terms of the used algorithms, treatment of radiation frequency dependence, the interaction of radiation with the gas, and the parallelization of methods for deployment on supercomputers. Broadly, the algorithms fall into two categories: (i) moment-based methods, encompassing the flux-limited diffusion approximation, M1 closure, and variable Eddington tensor methods, and (ii) methods directly solving the radiation transport equation, including forward and reverse ray tracing, characteristics-based methods, and Monte Carlo techniques. Beyond discussing advantages and disadvantages of these methods, the review also lists recent radiation hydrodynamic codes implemented the described methods.

\tiny
 \keyFont{ \section{Keywords:} star formation, interstellar medium, numerical techniques, radiation transport, astrophysical fluid dynamics} 
\end{abstract}

\section{Introduction}

Electromagnetic radiation plays a fundamental role in the process of star formation as it regulates the dynamics, thermodynamics, and chemistry of star-forming regions.  The interstellar radiation field (ISFR) heats the interstellar dust, dissociates molecules and affects the ionization balance, which is critical for the chemistry of the interstellar medium and the formation of molecular clouds. During the collapse of molecular cores, the dust absorbs the cooling radiation and once the medium becomes optically thick, the first proto-stellar core is formed. When the first stars are formed, their radiation exerts a profound influence on the surrounding gas and dust shaping the morphology of molecular clouds. This set of processes is called the radiative feedback as it regulates the rate and efficiency of star formation. Extreme ultraviolet photons from young, massive stars can ionize the surrounding gas, creating HII regions. The ionizing radiation also influences the chemistry of the gas, affecting the abundance of molecules necessary for the cooling and fragmentation of the cloud. Furthermore, radiation pressure from newly formed stars can counteract gravitational collapse, slow down the gas inflow and drive the winds and outflows.

The inclusion of radiation transport in star formation simulations is therefore desirable. However, it is also very hard due to both the high computational costs and the complexity of the processes that have to be taken into account. The radiation transport equation (RTE) in its general form reads
\begin{equation}
\frac{1}{c}\frac{\partial I_{\nu}}{\partial t} 
+ \mathbf{n}\cdot \nabla I_\nu =
- (\kappa_{\nu,a} + \kappa_{\nu,s}) \rho I_\nu + j_\nu\rho 
+ \frac{1}{4\pi} \kappa_{\nu,s}\rho \int_{\Omega} I_\nu d\Omega
\label{eq:rte}
\end{equation}
where the radiation field is represented by the specific intensity $I_\nu$. It is defined through the element of the radiation energy $dE_\nu$ flowing across an area element $da$ located at position $\mathbf{x}$ in time $dt$ in the solid angle $d\Omega$ about the direction $\mathbf{n}$ in the frequency interval $\nu + d\nu$ as
\begin{equation}
dE_\nu =  I_\nu(\mathbf{x}, \mathbf{n},t) \cos(\theta)\,d\nu\,da\,d\Omega\,dt
\label{eq:Inu}
\end{equation}
where $\theta$ is the angle that $\mathbf{n}$ makes with a normal to the area element $da$. The first term on the r.h.s. of Equation~\ref{eq:rte} represents the radiation losses due to absorption (described by the absorption opacity $\kappa_{\nu,a}$) and scattering (described by the scattering opacity $\kappa_{\nu,}$); $\rho$ is the mass density of the medium through which the radiation propagates. The second term on the r.h.s. of Equation~\ref{eq:rte} represents the emission of radiation describe by the emission coefficient $j_\nu$, the last r.h.s term represents contribution of the radiation scattered from other directions. Note that in the above equations, the specific intensity is expressed in the frame of coordinate system, however, emission coefficient and opacities are defined in the comoving frame.

The structure of Equations~\ref{eq:rte} and \ref{eq:Inu} reveals that the computational costs associated with solving them directly in their complete generality are exceedingly high. Specifically, in the case of three-dimensional problems, $I_\nu$ becomes a function of three components of $\mathbf{x}$, two components of $\mathbf{n}$, time $t$, and the frequency of the radiation $\nu$ (written as a subscript, as it is often grouped into several ranges known as wavebands) -- amounting to seven independent variables. This issue becomes particularly pronounced when implementing the radiation transport algorithm within a star formation simulation that demands high resolution and, consequently, a substantial number of computational points along the aforementioned variables. To address this, various computational techniques have been developed that solve only a portion of Equation~\ref{eq:rte} (e.g., incorporating only a subset of the processes it describes) or utilize diverse approximations, including those transforming it into a different (albeit related) equation that can be solved more computationally efficiently.

This paper reviews the most common techniques to solve the radiation transport equation within (magneto-)hydrodynamic simulations of star formation. The focus of this work is primarily on continuum radiation transport, rather than line transport (atomic or molecular). There is an extensive amount of work on solving radiation transport on static grids or as tools for synthetic observations, however, covering it is beyond the scope of this work. Additional information can be found e.g. in classical works on radiation hydrodynamics \citet{1984Mihalas, 2004Castor}, in a review of dust radiative transfer by \citet{2013Steinacker} and in a review of numerical methods in simulations of star formation by \citet{2019Teyssier}. The paper is organised as follows. Section \S\ref{sec:class} describes different way how radiation transport codes can be classified. Section \S\ref{sec:alg} reviews the most common radiation transport algorithms used in star formation, consisting of \S\ref{sec:mom} discussing the moment-based methods and \S\ref{sec:direct} describing the methods solving Equation~\ref{eq:rte} directly. Finally, Section \S\ref{sec:summary} summarizes this review.

\section{Classifications of the radiation transport treatments}
\label{sec:class}

There are several criteria based on which radiation transport codes can be classified. 

Firstly, the method can either explicitly include the first term of Equation~\ref{eq:rte} (partial time derivative) or assume the infinite speed of light and solve the RTE equation implicitly. The former approach, also known as the {\em evolutionary} method, is typically employed by moment-based methods. In some cases, the speed of light is artificially reduced to avoid excessively small time steps mandated by the stability conditions of an explicit solver -- a technique demonstrated to be useful in other fields as well \citep[e.g.][]{1999Cohen}. The implicit approach, alternatively termed the {\em instantaneous} method, is often utilized by codes based on the method of characteristics, ray tracing, and Monte Carlo methods.

Secondly and related to the above, the method can either directly solve Equation~\ref{eq:rte} (albeit with some approximations) or transform it into another type of equation. The former approach includes ray tracing, long and short characteristic methods, and Monte Carlo methods. While ray tracing and long characteristic methods are among the most accurate, they are also computationally expensive. The computational cost of ray tracing typically scales with the number of discrete radiation sources, making it particularly expensive for simulations with a high number of sources. In the latter approach, Equation~\ref{eq:rte} is commonly transformed into a hyperbolic system (M1 closure methods) or a parabolic system (flux-limited diffusion method).

Thirdly, various methods concentrate on distinct wavelengths of radiation, incorporating different physical processes related to the interaction between radiation and matter. In the simplest approach, termed {\em monochromatic}, all photons are assumed to possess a single (representative) frequency. A slightly more complex method is the {\em grey} approximation, where a specific constant spectrum -- often the black body spectrum -- is assumed for all radiation. In certain scenarios, it becomes necessary to categorize radiation based on its frequency into multiple groups that exhibit qualitative differences (e.g., where only some frequencies ionize a specific species). This approach is referred to as {\em multi-group} frequency treatment. Lastly, certain codes have the capability to encompass full frequency dependence by sampling the frequency with a designated number of bins.

An essential consideration in implicit radiation transport methods is whether the opacity at a specific location is depends on the radiation intensity at that point. A notable instance of this behavior is the treatment of ionizing (EUV) radiation using the on-the-spot approximation. In this approach, ionizing photons are effectively destroyed by case B recombinations, as they do not result in the replication of ionizing photons. The number of case B recombinations occurring in a given gas parcel needs to be distributed among all the rays passing through it, particularly when using ray tracing or the long characteristic method. Consequently, such methods necessitate iteration, leading to an increase in computational cost.

Finally, different methods pose varying challenges in terms of parallelization, with particular emphasis on the parallelization using domain decomposition suitable for distributed memory machines. Many hydrodynamic codes opt for this type of parallelization, as distributed memory machines offer extensive memory and processor core capabilities. Unfortunately, this parallelization scheme is also the most complex from a coding perspective. Radiation transport, being primarily a long-distance interaction problem, presents significant difficulties in parallelization. This is especially true for ray tracing, the long characteristics method, and Monte Carlo, where numerous calculations over considerable distances must be executed within a limited timeframe.

\section{Radiation transport algorithms}
\label{sec:alg}

\subsection{Moment-based methods}
\label{sec:mom}

Instead of the specific intensity $I_\nu$, the radiation field can be described by its moments obtained by integrating $I_\nu$ over all directions. The first three moments are defined as \citep[see e.g.][]{1991Shu}: 
\begin{equation}
\left(
\begin{array}{c}
     cE_\nu  \\
     \mathbf{F}_\nu \\
     c\mathbb{P}_\nu
\end{array}
\right)
= \oint 
\left(
\begin{array}{c}
     1  \\
     \mathbf{n} \\
     \mathbf{n\otimes n}
\end{array}
\right)
I_\nu d\Omega
\label{eq:radmoms}
\end{equation}
where $\mathbf{n}$ is a unit vector. The mean radiation energy density $E_\nu$, the radiation flux $\mathbf{F}_\nu$, and the radiation pressure tensor $\mathbb{P}_\nu$ are the zeroth, first and second moments of the radiation field, respectively. The radiation pressure is often expressed as $\mathbb{P}_\nu = \mathbb{D} E_\nu$ where $\mathbb{D}$ is the so called Eddington tensor.

Moments of Equation~\ref{eq:rte} can be obtained by multiplying it with a certain power of $\mathbf{n}$ and integrating it over all directions. The scattering is assumed to be isotropic. Then, the first two moment equations are
\begin{equation}
    \frac{\partial E_\nu}{\partial t} + \nabla\cdot \mathbf{F}_\nu = 4\pi\rho j_\nu - \rho\kappa_{\nu,a} c E_\nu
\label{eq:mom0}
\end{equation}
and
\begin{equation}
    \frac{1}{c}\frac{\partial \mathbf{F}_\nu}{\partial t} + c \nabla\cdot\mathbb{P}_\nu = -\rho\kappa_{\nu,a}\mathbf{F}_\nu \ .
\label{eq:mom1}
\end{equation}
In the above equations, the opacity and the emission coefficient are expressed in the frame of reference co-moving with the fluid (where they are typically isotropic), however, $E_\nu$, $\mathbf{F}_\nu$ and $\mathbb{P}_\nu$ are expressed in the coordinate system of the simulation (the {\em laboratory} frame). This problem is usually solved by either transforming Equations~\ref{eq:mom0} and \ref{eq:mom1} into the co-moving frame (CMF) or by transforming $\kappa_{\nu,a}$ and $j_\nu$ into the laboratory frame using the first order expansion ({\em mixed frame} formulation). The radiation moment equations in the co-moving frame have a form \citep[see][]{1984Mihalas}
\begin{equation}
 \frac{\partial E_\nu}{\partial t} + \nabla \cdot(\mathbf{u}E_\nu) + \nabla\cdot \mathbf{F}_\nu + \mathbb{P}:\mathbf{u} = 4\pi\rho j_\nu - \rho\kappa_{\nu,a} c E_\nu
\label{eq:mom0:cmf}
\end{equation}
and
\begin{equation}
    \frac{1}{c}\frac{\partial \mathbf{F}_\nu}{\partial t} + \frac{1}{c} \nabla\cdot(\mathbf{uF}_\nu) +  c \nabla\cdot\mathbb{P}_\nu = -\rho\kappa_{\nu,a}\mathbf{F}_\nu 
\label{eq:mom1:cmf}
\end{equation}
where $\mathbf{u}$ is the velocity of the gas. For details on the mixed frame formulation see e.g. \citet{1982Mihalas}. The advantage of the {\em mixed frame} approach is that the radiation transfer equations are significantly simpler and that it conserves the total energy, however, it is not suitable for computing for instance line opacities in the supersonic flow due to their rapid changes \citep{2022Moens}.

\subsubsection{Flux limited diffusion (FLD)}
\label{sec:fld}

The simplest moment-based method was derived as an extension of the radiation transport in a diffusion limit \citep{1978Minerbo,1981Levermore}. If the photon free mean path is small and the optical depth is high, the radiation is isotropic and the Eddington tensor has form $\mathbb{D} = \frac{1}{3}\mathbb{I}$ where $\mathbb{I}$ is the identity tensor. Assuming the radiation field is stationary, i.e. neglecting the first two terms in Equation~\ref{eq:mom1:cmf}, and considering $\nabla\cdot \mathbb{P}_\nu = \frac{1}{3} \nabla E_\nu$, the radiation flux is related to the radiation energy density as
\begin{equation}
    \mathbf{F}_\nu = -\frac{c}{3\kappa_\nu \rho} \nabla E_\nu
    \label{eq:fick}
\end{equation}
and the radiation transport problem reduces to the Fick's diffusion law. 

Relation~\ref{eq:fick} is not valid if the optical depth is low, because $\mathbf{F}_\nu$ can exceed the physical limit $\mathbf{F}_\mathrm{max} = c E_\nu$ there. Therefore, the FLD method introduces the so called {\em flux limiter} $\lambda$ and modifies Equation~\ref{eq:fick} in the following way
\begin{equation}
    \mathbf{F}_\nu = -\frac{\lambda(R) c}{\kappa_\nu \rho} \nabla E_\nu \ .
    \label{eq:fld:flux}
\end{equation}
The flux limiter $\lambda(R)$ is defined as a function of the normalized radiation energy density gradient
\begin{equation}
    R = \frac{|\nabla E_\nu|}{\kappa_\nu \rho E_\nu} \ ,
    \label{eq:fld:R}
\end{equation}
in such a way that it ensures Equation~\ref{eq:fld:flux} reduces to the Fick's law in the optically thick regime (i.e. $\lambda = 1/3$) and to the maximum allowed flux, $\mathbf{F}_\mathrm{max}$, in the optically thin regime (i.e. $\lambda = 1/R$). Various flux limiters fulfilling the above requirements have been suggested in the literature \citep[see e.g.][]{1978Minerbo, 1981Levermore, 1984Levermore, 2001Turner}. As an example we give the flux limiter of \citet{1981Levermore} in the form
\begin{equation}
  \lambda(R) = \frac{1}{R} \left( \mathrm{coth}(R) - \frac{1}{R}\right)
  \simeq \frac{2+r}{6 + 3R + R^2}
\end{equation}
and the corresponding approximate Eddington tensor
\begin{equation}
    \mathbb{D} = \frac{1}{2}(1-f)\mathbb{I} + \frac{1}{2}(3f-1)(\mathbf{n}\otimes\mathbf{n})
    \qquad \mathrm{where}\qquad
    f = \lambda + \lambda^2 R^2 \ .
\end{equation}

The FLD method then solves only the $0$-th moment equation in the form
\begin{equation}
  \frac{\partial E_{\nu}}{\partial t} + \nabla \cdot(\mathbf{u}E_{\nu}) + \nabla\cdot \frac{c\lambda(R)}{\kappa_\nu \rho} \nabla E_{\nu} + \mathbb{P}:\mathbf{u} = 4\pi\rho j_\nu - \rho\kappa_{\nu,a} c E_\nu 
\label{eq:fld}
\end{equation}
which is, from a mathematical point of view, a parabolic equation, and hence the standard parabolic solvers can be used.

The above equations are {\em monochromatic}, i.e. they are written for a single wavelength of the radiation. However, it is often practical to use the FLD method (and other methods as well) for radiation with a certain range of wavelengths. Then we speak about the {\em grey} radiation transport if there is a single range covering the broad wavelength range important for a given problem, or about the multi-group / multi-band radiation transport if there are several wavelength ranges. Additionally, the grey methods often use the Planck black body law, $B_\nu(T)$, as the source function, and then the $0$-th moment equation has form
\begin{equation}
  \frac{\partial E_{r}}{\partial t} + \nabla \cdot(\mathbf{u}E_{r}) + \nabla\cdot \frac{c\lambda}{\kappa_\mathrm{R}} \nabla E_{r} + \mathbb{P}:\mathbf{u} = \kappa_\mathrm{P} ( a_r T^4 - c E_{r} )
\label{eq:fldgrey}
\end{equation}
where $E_r \equiv \int E_\nu d\nu$ is the bolometric radiation energy density, $T$ is the gas/dust temperature, $a_r = 4\sigma_\mathrm{B}/c$ is the radiation constant, $\sigma_\mathrm{B}$ is the Stefan-Boltzmann constant, and $\kappa_\mathrm{P}$ and $\kappa_\mathrm{R}$ are Planck and Rosseland mean opacities defined as
\begin{equation}
\kappa_\mathrm{P} \equiv \frac{\int \kappa_\nu B_\nu(T) d\nu}{\int B_\nu(T) d\nu} 
\end{equation}
and
\begin{equation}
\frac{1}{\kappa_\mathrm{R}} \equiv \frac{\int \frac{1}{\kappa_\nu} \frac{\partial B_\nu(T)}{\partial T} d\nu}
{\int \frac{\partial B_\nu(T)}{\partial T} d\nu} \ .
\end{equation}

RHD codes using the FLD method are discussed below and summarized in Table~\ref{tab:fld} giving a reference to the code paper, code name, whether it uses co-moving or mixed frame of reference, whether it is parallel with the domain decomposition and some additional information. The first usage of the FLD method is probably by \citet{1973Alme} for the problem of  the X-ray emission resulting from the accretion of material onto a neutron star. In the field related to star formation, the grey FLD method was used by \citet{1989Kley,1990Bodenheimer,1999Klahr,2001Boss} for simulations of accretion discs around young low mass stars. It yielded reasonably correct temperatures in the optically thick and optically thing regions, however, it showed deviations from the correct solution in the transition regions \citep{2007Boley}. The treatment of radiation in such models has been significantly improved by \citet{2010Kuiper} who developed a hybrid scheme combining the grey FLD method with the frequency dependent ray tracing, and implemented it to the MHD code \textsc{Pluto}. They at first calculate the stellar radiation up to the first absorption by a simple ray tracing algorithm and subsequently, they use FLD to calculate the radiation  re-emitted by the dust. This approach turned out to be very successful and many other authors implemented this or similar schemes into their codes. \citet{2013Bitsch} used it with the \textsc{Nirvana} code to study the influence of opacity and stellar irradiation on the disc structure and on the migration of planets. \citet{2013Flock} and \citet{2013Kolb} created two independent implementations of this hybrid method for the \textsc{Pluto} code. \citet{2014Klassen} generalized this method for multiple sources and arbitrary geometry wrote it as a module for the MHD code \textsc{Flash}. Similarly, \citet{2015Ramsey} implemented a generalized version of this method for the \textsc{AZEuS} code -- a \textsc{Zeus} family AMR code with a fully staggered mesh.

The simplicity of the FLD method allowed it to be implemented into many general purpose (magneto-)hydrodynamic codes used in astrophysics. The widely used \textsc{Zeus2d} code was originally released with the radiation module based on the tensor variable Eddington factor, however, the FLD method was written for it by \citet{2001Turner}. The \textsc{Orion} code used for many simulations of star formation, including the feedback from massive stars, was developed with FLD method being part of it \citep{2007Krumholz}. \citet{2008Gittings} developed the AMR code \textsc{Rage} including the grey FLD, for which they introduce a new technique of solving the diffusion and material energy equations, allowing larger time steps and more robust behavior. \citet{2011Commercon} implemented FLD into the AMR code \textsc{Ramses}, and \citet{2014Commercon} added adaptive time stepping and the implicit time integration. The block AMR code \textsc{Crash} \citep{2011vanderHolst} was developed with grey FLD where electrons and ions are allowed to have different temperatures. The grey FLD method in its mixed-frame formulation is also part of the AMR code \textsc{Castro} \citep{2011Zhang}, one of the first 3D RHD codes using the unsplit PPM solver. \textsc{Trhd} \citep{2015Sijoy} is two-dimensional unstructured mesh Lagrangian code using the grey FLD with three temperature approach (ions, electrons and radiation are allowed to have different temperatures). The \textsc{Flash} code \citep{2000Fryxell}, a computational framework consisting of modules implementing various physical processes including hydrodynamics and radiation transport, was provided by the FLD module implemented by \citet{2019Chatzopoulos} and tested on 1D simulations of supernova Ia explosions. The general fluid dynamics code \textsc{Gizmo} \citep{2019Hopkins} includes several radiation transport solvers including the one based on FLD. Recently, the grey FLD method was implemented into the finite-volume MHD code \textsc{Mpi-Amrvac} \citep{2022Moens}.

The FLD method was implemented also into the Smoothed Particle Hydrodynamics (SPH) codes. Basic methods for calculating radiative heat diffusion were described in seminal works on SPH by \citet{1977Lucy} and \citet{1985Brookshaw}. The full grey FLD method was implemented into the SPH by \citet{2004Whitehouse} and \citet{2005Whitehouse}. Later, this code originating from code \textsc{SphNG} \citep{1990Benz} was used by \citet{2009Price,2015Bate,2019Bate} and many other works for simulating star forming clouds, where the FLD method can be efficiently used to follow collapsing pre-stellar cores with high optical depths. \citet{2007Mayer} used implemented FLD into SPH code \textsc{Gasoline} and used it to study the gravitational instability in protoplanetary discs. \citet{2006Fryer} developed code \textsc{Snsph} including the FLD radiation transport and used it for simulating supernova explosions, where FLD can also be used efficiently due to high optical depths. Another implementations of FLD into SPH codes are by \citet{2006Viau} and recently by \citet{2021Bassett} (code \textsc{Spheral}).

\begin{table}[]
  {\small
  \centering
  \begin{tabular}{|l|l|l|l|l|}
    \hline
    Reference                   & Name/code & frame & DD & Note \\       
    \hline
    \citet{1989Kley}            & -                    & CMF   & N & 2D axisym, protoplanetary discs\\
    \citet{1990Bodenheimer}     & -                    & CMF   & N & 2D HD code by \citet{1985Rozyczka} \\
    \citet{1999Klahr}           & \textsc{Tramp}       & CMF   & N & \\
    \citet{2001Boss}            & -                    & CMF   & N & 3D disc forming giant planet\\
    \citet{2001Turner}          & \textsc{Zeus2d}      & CMF   & N & \\
    \citet{2006Fryer}           & \textsc{Snsph}       & CMF   & Y & SPH code, neutrino diffusion \\
    \citet{2006Viau}            & -                    & CMF   & N & SPH code \\
    \citet{2007Krumholz}        & \textsc{Orion}       & mixed & Y & \\
    \citet{2007Mayer}           & \textsc{Gasoline}    & CMF   & Y & \\
    \citet{2008Gittings}        & \textsc{Rage}        & CMF   & Y & rad-matter coupling $\rightarrow$ larger dt \\
    \citet{2010Kuiper}          & \textsc{Pluto}       & negl. & Y & + $\nu$-dep ray tr. \\
    \citet{2011Commercon}       & \textsc{Ramses}      & CMF   & Y & \\
    \citet{2011vanderHolst}     & \textsc{Crash}       & CMF   & Y & multiple frequency groups\\
    \citet{2011Zhang}           & \textsc{Castro}      & mixed & Y & unsplit PPM solver \\
    \citet{2013Bitsch}          & \textsc{Nirvana}     & negl. & N & + $\nu$-dep ray tr. in rad. dir. \\
    \citet{2013Flock}           & \textsc{Pluto}       & negl. & Y & + $\nu$-dep ray tr. \\
    \citet{2013Kolb}            & \textsc{Pluto}       & negl. & Y & + $\nu$-dep ray tr. \\
    \citet{2014Klassen}         & \textsc{Flash}       & CMF   & Y & + $\nu$-dep ray tr., multiple src. \\
    \citet{2015Bate}            & \textsc{sphNG}       & CMF   & Y & + TreeCol \citep{2012Clark} \\
    \citet{2015Sijoy}           & \textsc{Trhd}        & CMF   & N & 3-temp, 2D unstructured-mesh \\
    \citet{2015Ramsey}          & \textsc{AZEuS}       & CMF   & N & + $\nu$-dep ray tr. \\
    \citet{2019Chatzopoulos}    & \textsc{Flash}       & mixed & Y & opacities from OPAL   \\
    \citet{2019Hopkins}         & \textsc{Gizmo}       & mixed/CMF & Y & multiple frequency groups \\
    \citet{2021Bassett}         & \textsc{Spheral}     & CMF   & Y & SPH code \\
    \citet{2022Moens}           & \textsc{Mpi-Amrvac}  & CMF   & Y & finite-volume MHD code \\
    \hline
  \end{tabular}
  \caption{Radiation transport codes using the flux limited diffusion (FLD) approximation. The first column gives the reference to the work describing the code or algorithm; the second column give the name of the radiation transport code or RHD code (if they exist); the third column specifies the frame of reference in which the radiation energy equation is formulated (mixed, co-moving -- CMF, or radiation energy advection terms are neglected); the fourth column tells whether the code is parallel using the domain decomposition; and the last column gives eventually an additional information.}
  \label{tab:fld}
  }
\end{table}

\subsubsection{M1 closure}
\label{sec:m1}

A more complex and accurate two-moments method is the M1 closure \citep{1999Dubroca.329..915D}. It assumes that the Eddington tensor is axially symmetric around the flux vector and hence can be expressed as
\begin{equation}
    \mathbb{D} = \frac{1-\chi}{2} \mathbb{I} + \frac{3\chi-1}{2} \mathbf{n}\otimes \mathbf{n}
\end{equation}
where $\chi$ is a scalar function called the {\em Eddington factor}. The M1 closure method uses one of its simplest forms discussed in \citet{1984Levermore}, corresponding to the Lorentz boosted isotropic distribution of the radiation, 
\begin{equation}
    \chi = \frac{3 + 4|\mathbf{f}|^2}{5 + 2\sqrt{4 - 3|\mathbf{f}|^2}}
\end{equation}
where $\mathbf{f} = |\mathbf{F}_\nu|/cE_\nu$ is the reduced flux. Then, the method solves Equations~(\ref{eq:mom0:cmf}) and (\ref{eq:mom1:cmf}) or their equivalent in the mixed frame approach.

The advantage of the M1 closure is that it captures correctly the two extremes of the radiation field: (i) the free streaming (or optically thin) limit with $|\mathbf{f}| = 1$ and $\chi = 1$, and (ii) the diffusion (or optically thick) limit with $|\mathbf{f}| = 0$ and $\chi = 1/3$. The ability to maintain the original direction of the free streaming radiation is a pronounced improvement in comparison to the FLD method leading to the artificial diffusion of the free streaming radiation. On the other hand, the assumed simple form of the radiation pressure tensor cannot correctly describe more complex radiation fields as for instance the radiation from multiple sources in which case the nonphysical interaction would occur.

A specific feature of the M1 closure making is suitable for the implementation into (magneto-)hydrodynamic codes is that the system of Equations~\ref{eq:mom0:cmf} and \ref{eq:mom1:cmf} is hyperbolic and hence it can be solved with the same algorithms as the equations of the fluid dynamics. Table~\ref{tab:m1} gives a list of star formation and related fields codes implementing the M1 closure method. The first such code was \textsc{Heracles} \citep{2007Gonzalez} designed to study radiative shocks (and compare the results with laboratory experiments), jets of young stars, formation of pre-stellar cores and protoplanetary disks. This method was implemented into the cosmic reionisation code \textsc{Aton} \citep{2008Aubert}, and later ported to the general RHD code \textsc{Ramses} \citep{2013Rosdahl,2015Rosdahl}. \citet{2013Sadowski} included the M1 closure method into the general relativistic code \textsc{Koral} formulating the RHD equations in the covariant form. \citet{2013Skinner} describe the M1 closure module for the \textsc{Athena} code, in which they use the reduced speed of light approximation to improve the performance. The general fluid dynamics code \textsc{Gizmo} includes the M1 closure as one of its radiation transport solvers. \citet{2019Kannan} describe the radiation transport module for the unstructured moving mesh code \textsc{Arepo}, based on a new higher order implementation of the M1 closure method. This method is also included into code \textsc{Fornax} \citep{2019Skinner} primarily intended for core collapse supernova simulations, general fuild dynamics code \textsc{Pluto} \citep{2019MelonFuksman}, and hybrid OpenMP/MPI/GPU parallel code \textsc{Ark-rt} \citep{2021Bloch}. \citet{2021Chan} present module \textsc{Sph-m1rt} for task-based parallel SPH code \textsc{Swift}.

\begin{table}[]
  {\tiny
  \centering
  \begin{tabular}{|l|l|l|l|l|}
    \hline
    Reference            & Name/code     & frame & DD & Note \\       
    \hline
    \citet{2007Gonzalez}     & \textsc{Heracles}      & CMF       & Y &  \\
    \citet{2013Rosdahl}      & \textsc{Ramses-rt}     & CMF       & Y & from ATON \citep{2008Aubert} \\
    \citet{2013Sadowski}     & \textsc{Koral}         & covar.    & N & general relativistic MHD code\\
    \citet{2013Skinner}      & \textsc{Athena}        & mixed     & Y & reduced speed of light approx. \\
    \citet{2019Hopkins}      & \textsc{Gizmo}         & CMF/mixed & Y & multiple frequency groups \\
    \citet{2019Kannan}       & \textsc{Arepo}         & CMF       & Y & higher order scheme of M1 \\
    \citet{2019MelonFuksman} & \textsc{Pluto}         & mixed     & Y &  \\
    \citet{2019Skinner}      & \textsc{Fornax}        & CMF       & Y & photon and neutrino rad. fields \\
    \citet{2021Bloch}        & \textsc{Ark-rt}        & CMF       & Y & hybrid OpenMP/MPI/GPU parallel \\
    \citet{2021Chan}         & \textsc{Sph-m1rt}      & CMF       & Y & in SPH code SWIFT \\
    \hline
  \end{tabular}
  \caption{Radiation transport codes using the M1 closure method. The meaning of columns is the same as in Table.~\ref{tab:fld}.}
  \label{tab:m1}
  }
\end{table}


\subsubsection{Variable Eddington Tensor methods}
\label{sec:vet}

Another approach to solve moment Equations~\ref{eq:mom0} and \ref{eq:mom1}, as opposed to the closure assuming an approximate form of the Eddington tensor, is to calculate the Eddington tensor self consistently by directly solving the radiation transport Equation~\ref{eq:rte}. Methods using it are referred to as {\em Variable Eddington Tensor} (VET) methods, and the Eddington tensor, $\mathbb{D}$, is calculated by one of the methods described in \S\ref{sec:direct} (ray tracing, long or short characteristics, or Monte Carlo). Subsequently, $\mathbb{D}$ is combined with the radiation moment equations to calculate the radiation energies and fluxes.

The VET method was developed for 1D calculations of plane-parallel stellar atmospheres by \citet{1970Auer}, however, the idea of iterating the ratio of second and zeroth radiation moments dates back to \citet{1926Eddington}. The first multi-dimensional astrophysical code using this method was \textsc{Zeus2d} \citep{1992Stone}, where $\mathbb{D}$ was calculated by the short characteristics. The algorithm was later updated in code \textsc{Zeus-mp} \citep{2003Hayes} by adding more methods to calculate $\mathbb{D}$ and by separating the radiation and gas energy densities. The 1D RHD code \textsc{Titan} \citep{1994Gehmeyr} uses VET and calculates $\mathbb{D}$ by ray tracing. \citet{2001Gnedin} developed an algorithm based on the {\em Optically Thin Variable Eddington Tensor} (OTVET) approximation where $\mathbb{D}$ is calculated in the optically thin regime. This is done by integrating contributions of radiation sources  with attenuation factor $1/r^2$ over the computational domain taking advantage of the solver used for the self-gravity. They implemented this algorithm into the Soften Lagrangian Hydrodynamic code (SLH) following all physical quantities on a moving deformed mesh. This algorithm was later implemented by \citet{2009Petkova} into the SPH code \textsc{Gadget}. \citet{2010Sekora} developed a hybrid Godunov method based on VET with user defined $\mathbb{D}$ and implemented it into 1D RHD code \textsc{Nike}. This method was later implemented into code \textsc{Athena} by \citet{2012Jiang} where $\mathbb{D}$ is calculated by the short characteristics method described by \citet{2012Davis}. The general relativistic radiation magnetohydrodynamic (GR-RMHD) code \textsc{Inazuma} \citep{2020Asahina} also uses the VET based method to treat the radiation, and it calculates $\mathbb{D}$ integrating the local radiation intensity over the angular directions. Recently, the RHD code \textsc{Quokka} \citep{2022Wibking} implements the general two-moment radiation transport method. The code uses the block structured adaptive mesh refinement handled by the AmREX library and it is parallelized for both MPI and usage on graphic processing units (GPUs) using the CUDA library. Its radiation solver is written to use a general Eddington tensor, the default option is the M1 closure.

\begin{table}[]
  {\small
  \centering
  \begin{tabular}{|l|l|l|l|l|}
    \hline
    Reference                    & Name/code    & frame & DD & Note \\       
    \hline
    \citet{1992Stone}            &  \textsc{Zeus2d}      & CMF   & N & $\mathbb{D}$ cmp. by short chars. \\
    \citet{1994Gehmeyr}          &  \textsc{Titan}       & CMF   & N & 1D code; $\mathbb{D}$ cmp. by ray tracing\\
    \citet{2001Gnedin}           &  \textsc{Otvet}       & Lag   & N & $\mathbb{D}$ cmp. in optically thin approx\\
    \citet{2010Sekora}           &  \textsc{Nike}        & Lag   & N & hybrid Godunov RHD, $\mathbb{D}$ user defined\\
    \citet{2012Jiang}            &  \textsc{Athena}      & mixed & Y & $\mathbb{D}$ cmp. by short chars.\\
    \citet{2020Asahina}          &  \textsc{Inazuma}     & CMF   & N & $\mathbb{D}$ cmp. by integrating intensity \\
    \citet{2022Wibking}          &  \textsc{Quokka}      & mixed & Y & GPU, $\mathbb{D}$ general, M1 by default \\
    \hline
  \end{tabular}
  \caption{Radiation transport codes using the Variable Eddington Tensor (VET) approach. The meaning of columns is the same as
in Table.~\ref{tab:fld}}
  \label{tab:vet}
  }
\end{table}


\subsection{Methods solving RTE directly}
\label{sec:direct}

Table~\ref{tab:rt} lists codes that employ the direct solution of the RTE: forward and reverse ray tracing, methods of characteristics, and Monte Carlo methods.

\subsubsection{Ray tracing}
\label{sec:rayt}

The most straightforward method to solve the radiation transport Equation~\ref{eq:rte} is to follow the radiation from its sources along the lines of its propagation, a technique called {\em ray tracing}. It is widely used in many other fields, in particular in the computer image generation. The original idea of ray tracing dates back to 16th century to Albrecht D\"urer who described multiple techniques for projecting 3D scenes onto an image plane \citep{1990Hofmann}.


As previously mentioned, the primary challenge in directly solving Equation~\ref{eq:rte} is its high dimensionality, resulting in high computational costs. Therefore, a key characteristic of a ray tracing algorithm is a set of approximations designed to reduce the complexity of the problem. A crucial aspect is the method by which the radiation field is discretized in both space and its propagation directions. The simplest approach, employed by many early and recent codes, involves calculating the radiation on the same grid as the gas dynamics. This confines the direction of radiation propagation to the coordinate axes. Examples of this approach include \citet{1986Wolfire} in a 1D spherical grid, \citet{1994Murray,1996GarciaSegura} in a 2D spherical grid, and numerous other works, the review of which exceeds the scope of this study. This method is particularly well-suited for treating radiation emanating from a single source (e.g., a central star) located at the center of the coordinate system. It can be effectively combined with other methods that describe the diffuse radiation component (e.g., FLD, as mentioned in Section~\ref{sec:fld} and Table~\ref{tab:fld}).

In many cases, confining the radiation propagation solely along the hydrocode grid axes is overly restrictive. Therefore, general radiation transport codes typically employ an independent method to define rays, making ray tracing algorithms particularly suitable for implementation into unstructured grid or smooth particle hydrodynamic codes. One approach is to cast rays from a source in spherical coordinates, associating each ray with a spherical segment of the same size \citep{1999Abel,1999Razoumov}. This method was later refined in the HEAlPix library \citep{2005Gorski}, which has become a de facto standard. HEALPix facilitates the tessellation of the sphere surface into equal area regions in a hierarchical manner, a feature utilized by adaptive ray tracing algorithms that split rays in regions requiring finer angular resolution \citep{2002Abel}. This approach has proven highly successful and has been employed by various authors, including \citet{2008Pawlik} (\textsc{Traphic} code linked to SPH code \textsc{Gadget2}), \citet{2009Bisbas} (SPH code \textsc{Seren}), \citet{2011Wise} (\textsc{Moray} code linked to AMR code \textsc{Enzo}), \citet{2015Baczynski} (\textsc{Fervent} code within the AMR code \textsc{Flash}), and others.

Depending on the number of cast rays, ray tracing algorithms can achieve high accuracy, although with correspondingly higher computational costs that typically scale with the number of radiation sources. As a result, they are occasionally employed in conjunction with other methods where ray tracing specifically handles radiation from a relatively low number of discrete sources, such as massive stars. An example is the \textsc{Harm$^2$} code \citep{2017Rosen}, which calculates ionizing stellar radiation using adaptive ray tracing and the diffuse component using the FLD method in the ORION code. Similarly, the \textsc{Torus-3dpdr} code \citep{2015BisbasA} utilizes adaptive ray tracing for FUV radiation and the Monte Carlo method for the ionizing and diffuse components.

A slightly different approach to spatially adaptive ray tracing was used by \citet{2007Dale} who integrated the RTE on the line-of-sight between the source and the target particle using evaluation points defined by SPH particles close to the line-of-sight selected using method by \citet{2000KesselDeynet}.

\citet{2006Ritzerveld} developed algorithm SimpleX to solve Equation~\ref{eq:rte} on an unstructured grid calculated from the properties of the medium in which the radiation propagates, using the Delaunay triangulation \citep{1934Delaunay}. The computational costs of this method do not depend on the number of sources, similarly to moment-based methods or some reverse ray tracing schemes. It is typically less diffusive that moment based methods and it is more suitable for parallelisation with domain decomposition than most of the ray tracing methods. It is particularly suitable for being used with SPH or unstructured grid codes, \citet{2014Clementel} implemented it into an SPH code. The SimpleX algorithm was improved and parallelized by \citet{2010Paardekooper} and implemented as code \textsc{Sprai} integrated into code \textsc{Arepo} by \citet{2018Jaura}.

\subsubsection{Characteristics-based methods}
\label{sec:chars}

A specific way of ray tracing are methods of characteristics, which solve the Radiative Transfer Equation (RTE) along predefined rays (characteristics) covering the entire computational domain rather than casting rays from discrete sources. In this manner, characteristics-based methods can describe the diffuse radiation component. They can be categorized into {\em long}, {\em short} and {\em hybrid} characteristics based on the end points of the rays. Long characteristics extend across the entire computational domain, with their separation comparable to the size of the computational elements and defined in specific directions. While long characteristics are among the most accurate methods, they are also the most expensive, depending on the number of rays. Another disadvantage, applicable to some extent to all ray tracing methods, is that they are difficult to get parallelized within codes that use domain decomposition on distributed memory machines. This challenge arises because rays crossing multiple domains typically represent a substantial amount of information that needs to be communicated over the (relatively) slow network. An example of a code utilizing long characteristics is \textsc{Lampray} \citep{2018Frostholm}, implemented into the \textsc{Ramses} code. \textsc{Lampray} calculates both diffuse and discrete sources radiation on an adaptive oct-tree mesh, allowing the definition of several frequency bins for both ionizing and non-ionizing radiation. It is coupled with the non-equilibrium chemistry code \textsc{Krome} \citep{2014Grassi}, providing the chemical networks to evolve species abundances.

The short characteristics methods create rays only among the neighbour cells (or generally computational elements) and interpolate the radiation intensity at the cell borders. This makes this class of methods more diffusive, but also computationally cheaper and much easier to parallelize with domain decomposition codes. Short characteristics are used e.g. by code \textsc{C$^2$ray} \citep{2006Mellema},  code {\sc Athena} \citep{2012Davis}, or code \textsc{Pion} \citep{2021Mackey}. They have been also used together with the moment-based VET method to calculate the local radiation pressure tensor $\mathbb{D}$ (see \S\ref{sec:vet}).

\citet{2006Rijkhorst} developed the hybrid characteristics method, combining the advantages of both long and short characteristics, particularly in reducing the necessary parallel communication. This method is implemented within the block-based daptive AMR code \textsc{Flash}. It divides the radiation field into two components: (i) the local component within the blocks, calculated using the long characteristics method, and (ii) the global component, transporting radiation among blocks (and parallel sub-domains) using interpolation similar to the short characteristics method. This algorithm was further enhanced by \citet{2010Peters} and later re-implemented by \citet{2016Buntemeyer} to also handle diffuse radiation and operate on GPU-accelerated architectures.

\citet{2021Jiang} developed an implicit radiation transport algorithm for MHD code \textsc{Athena++}. It represents the radiation by its specific intensities along discrete rays, and evolves them using a conservative finite volume method. It avoids using a closure as in moment-based methods and the related artificial diffusion. On the other hand, it solves time-dependent radiation transport equation, contrary to majority of ray or characteristics based methods. The original implementation was grey, however, \citet{2022Jiang} extended it to include the frequency dependence using the multi-group approach.

\subsubsection{Reverse ray tracing}
\label{sec:revrt}

An alternative approach is the so-called {\em reverse ray tracing}. In contrast to the "standard" forward ray tracing, which casts rays from the source outward, reverse ray tracing schemes cast rays from the target point where the radiation interacts with the gas (e.g., grid cell or SPH particle) and transport the radiation inward to the target point. This approach offers several advantages, including the highest density of rays (i.e., the highest spatial resolution) near the location where radiation interacts with the gas. Another advantage is its suitability for handling external diffuse radiation. \citet{2010Hasegawa} implemented this algorithm into the SPH code \textsc{Start}, incorporating tree-based acceleration to group distant sources into oct-tree nodes. A grid-based version of this algorithm was developed by \citet{2012Okamoto} for the code \textsc{Argot}. \citet{2013Altay} created the reverse ray tracing code \textsc{Urchin} and integrated it into an SPH code focused on cosmological simulations. Another implementation of this approach is \textsc{Trevr} for the SPH code \textsc{Gasoline} by \citet{2019Grond}, who introduced adaptive refinement along rays and directionally dependent absorption coefficients for tree nodes. \citet{2021Wunsch} developed the code \textsc{Treeray}, implementing tree-accelerated reverse ray tracing for the Adaptive Mesh Refinement (AMR) code \textsc{Flash}. It utilizes a fast parallel tree solver described in \citet{2018Wunsch} and is coupled with the chemical network developed by \citet{2007Glover}. Later, it was extended by \citet{2023Klepitko}, who added diffuse grey infrared radiation and related radiation pressure, and by \citet{2023Gaches}, who added frequency-dependent X-ray radiation. \textsc{Treeray} has been used in the SILCC project \citep[][and follow-up papers]{2015Walch} using RMHD simulations to model the full cycle of molecular clouds in a part of a galactic disc.


\subsubsection{Monte Carlo methods}
\label{sec:mc}

Monte Carlo Radiation Transport (MCRT) algorithms belong to a broad category of statistical sampling methods that share this name. They simulate the movement of individual photons through a medium by stochastically sampling random events, including interactions with matter. Given that tracking individual photons is prohibitively expensive, MCRT algorithms introduce "photon packets" that carry information about the position, direction, frequency distribution, etc., of the group of photons they represent. This approach closely mirrors the reality of propagating radiation, making it highly versatile, and facilitating the inclusion of various complex processes such as scattering.

However, the statistical nature of this method introduces Poisson noise, and maintaining it below a required level often necessitates a very large number of photon packets. Consequently, MCRT codes have historically mostly been used for static problems (e.g., the code \textsc{Mocassin} by \citealt{2003Ercolano}, the code \textsc{Hyperion} by \citealt{2011Robitaille}, or the code \textsc{Radmc3D} by \citealt{2012Dullemond}). Due to the potential long distances photon packets can travel, Monte Carlo methods are relatively challenging to parallelize using domain decomposition. On the other hand, thread-based parallelization on shared memory architectures is typically straightforward.

As this method does not rely on any grid, it is natural to use it with SPH or unstructured grid codes. An algorithm to incorporate MCRT into SPH codes was proposed by \citet{1999Lucy}. A Monte Carlo-based ray-tracing algorithm was developed by \citet{2008Altay} and implemented into the publicly available code \textsc{Sphray}, primarily focused on cosmological simulations. Another implementation of MCRT into an SPH code was done by \citet{2009Nayakshin}, who used it to calculate the radiation pressure force. \citet{2015Roth} developed a 1D MCRT code and coupled it with both Eulerian and Lagrangian hydrodynamic codes. The first star formation simulations with MCRT were performed by \citet{2016Lomax}, who developed the code \textsc{Spamcart} coupled with SPH hydrodynamics. They used it to calculate synthetic intensity maps and spectra of an embedded protostellar multiple system. \citet{2018Vandenbroucke} developed the radiation hydrodynamic code \textsc{CMacIonize}, which works on an unstructured moving grid and is based on the Monte Carlo method calculating $\nu$-dependent transport of ionizing photons. The code is parallelized using a task-based scheme and has an option to work in a distributed memory configuration. Code \textsc{Torus}, developed by \citet{2019Harries}, is an MCRT that includes native AMR grid-based hydrodynamics and can also be coupled with the SPH code \textsc{Sphng}. It implements numerous microphysics modules, including atomic and molecular line transport in moving media, dust radiative equilibrium, photoionization equilibrium, and time-dependent radiative transfer, and has been used e.g. for modelling massive stars feedback in turbulent clouds \citep{2018Ali}. It is also designed for postprocessing the output of several other (magneto-)hydrodynamic codes. Recently, MCRT has also been implemented into the unstructured moving mesh code \textsc{Arepo} by \citet{2020Aaron}.

\begin{table}[]
  {\tiny
  \centering
  \begin{tabular}{|l|l|l|l|l|l|}
    \hline
    Reference                 & Name/code           & $t$-dep & DD & MF & Note \\       
    \hline
    \citet{2002Abel}          & -                            & N       & N  & ion. & gen., adaptive RT with HealPix \\
    \citet{2006Mellema}       & \textsc{C$^2$ray}            & N       & N  & ion. & grid, short chars.\\
    \citet{2006Rijkhorst}     & \textsc{Flash}               & N       & Y  & grey & grid, hybrid characteristics \\
    \citet{2007Dale}          & \textsc{Sphng}               & N       & Y  & ion. & SPH, adaptive RT \\
    \citet{2008Altay}         & \textsc{Sphray}              & N       & N  & $\nu$-dep & SPH, Monte Carlo \\
    \citet{2008Pawlik}        & \textsc{Traphic/Gadget2}     & Y       & Y  & $\nu$-dep & SPH, photon packets, cones\\
    \citet{2009Bisbas}        & \textsc{Seren}               & N       & N  & ion. & SPH, adaptive RT \\
    \citet{2009Nayakshin}     &  -                           & Y       & N  & grey & SPH, rad. pressure\\
    \citet{2010Hasegawa}      & \textsc{Start}               & N       & N  & ion. & SPH, reverse RT, tree accel. \\
    \citet{2011Wise}          & \textsc{Moray/Enzo}          & Y       & Y  & MG   & grid, adaptive RT\\
    \citet{2012Okamoto}       & \textsc{Argot}               & N       & Y  & ion. & grid-based version of START \\
    \citet{2012Clark}         & \textsc{Treecol}             & N       &    &      & reverse RT / column densities \\
    \citet{2012Davis}         & \textsc{Athena}              & N       & Y  & grey & grid, short chars. \\
    \citet{2013Altay}         & \textsc{Urchin}              & N       & N  & $\nu$-dep & SPH, reverse RT\\
    \citet{2015Baczynski}     & \textsc{Fervent/Flash}       & N       & Y  & MG        & grid, forward RT\\
    \citet{2015BisbasA}       & \textsc{Torus-3Dpdr}         & N       & Y  & $\nu$-dep & SPH, adp. RT (FUV) + MC (ion.\&diff.)\\
    \citet{2015Roth}          & -                            & Y       & N  & $\nu$-dep & grid, 1D Eulerian and Lagrangian \\
    \citet{2016Buntemeyer}    & \textsc{Flash}               & N       & Y  & grey & grid, hybrid chars on adaptive mesh, GPU\\
    \citet{2016Lomax}         & \textsc{Spamcart}            & Y       & N  & $\nu$-dep & SPH, first MCRT SF sims.\\
    \citet{2017Rosen}         & \textsc{Harm$^2$/Orion}      & N+Y     & Y  & $\nu$-dep & grid, forward RT ($\nu$-dep) + grey FLD\\
    \citet{2018Frostholm}     & \textsc{Lampray/Ramses}      & N       & Y  & MG & grid, long chars., coupled to KROME \\
    \citet{2018Jaura}         & \textsc{Sprai/Arepo}         & N       & Y  & ion. & unstruct. grid, SimpleX2 alg. \\
    \citet{2018Vandenbroucke} & \textsc{CMacIonize}          & N       & Y  & $\nu$-dep & unstruct. grid, Monte Carlo \\
    \citet{2019Grond}         & \textsc{Trevr/Gasoline}      & N       & Y  & ion. & SPH, reverse RT, tree\\
    \citet{2019Harries}       & \textsc{Torus}               & Y       & Y  & $\nu$-dep & grid, Monte Carlo \\
    \citet{2020Aaron}         & \textsc{Arepo-MCRT}          & Y       & Y  & grey & unstruct. grid, Monte Carlo \\
    \citet{2021Jiang}         & \textsc{Athena++}            & Y       & Y  & MG   & grid, rays + FVM, implicit \\
    \citet{2021Mackey}        & \textsc{Pion}                & N       & Y  & MG   & grid, short chars. \\
    \citet{2021Wunsch}        & \textsc{TreeRay/Flash}       & N       & Y  & MG & grid, reverse RT, tree \\
    \hline
  \end{tabular}
  \caption{Radiation transport codes using ray tracing and methods of characteristics. The first column gives the reference to the work describing the code or algorithm; the second column give the name of the radiation transport code or RHD code (if they exist); the third column whether the code solves time-dependent RTE; the fourth column tells whether the code is parallel using the domain decomposition; the fifth column gives how the code deals with the frequency dependence (grey, ion. - a single band of ionising photons, MG - multi-group approach, $\nu$-dep - full frequency dependence); and the last column gives eventually an additional information.
  }
  \label{tab:rt}
  }
\end{table}

\subsection{Other approximations}
\label{sec:other}

Many studies conducting hydrodynamic simulations of star formation further simplify the radiation transport problem. For example, \citet{2017VazquezSemadeni} include stellar ionizing radiation feedback by calculating the Str\"omgren sphere radius from the EUV photon production rate of a star and the characteristic density obtained as the geometric mean of the density at the star and at the target cell. Subsequently, they set the gas temperature within the sphere to $10^4$\,K.

A specific case involves non-ionizing diffuse ambient interstellar radiation fields, crucial for the physics of molecular clouds as they provide heating and affect molecular gas chemistry. In molecular clouds, this radiation intensity strongly depends on shielding, i.e., the optical depth between a given location and the edge of the cloud. This consideration led \citet{2012Clark} to develop the \textsc{Treecol} algorithm, implemented into the code \textsc{Gadget2}. It efficiently calculates optical depths in all directions (discretized by HealPix) for each SPH particle using a tree, enabling the determination of heating and dissociation rates. This algorithm was later implemented by \citet{2014Valvidia} into \textsc{Ramses} and by \citet{2018Wunsch} into \textsc{Flash}.

A different approach to solving a very similar problem was suggested by \citet{2007Stamatellos}, who estimate the mean optical depth from local density, temperature, and gravitational potential using a set of precalculated models of collapsing clouds. This method provides reasonably good results for almost zero computational costs.

The simplest treatment of radiation is to modify the gas equation of state to calculate the equilibrium gas temperature from local gas density, thus accounting for the most probable result of the heating-cooling balance. A popular choice is the barotropic equation of state \citep[e.g.,][]{1994Bonnell,1995Whitworth,1998Bate}, a simple analytical formula designed to mimic the thermodynamics of spherically symmetric collapse of a single, isolated protostar.

\section{Summary}
\label{sec:summary}

This work reviews methods to solve radiation transport equation (RTE) in simulations of star formation and related fields. According to the algorithm, they can be roughly divided into two groups: moment-based methods and methods solving the RTE directly.

Moment-based methods approximate the radiation specific intensity by integrating it over all directions and expressing it in terms of several moments. The order of the moments in the sequence closure determines the level of approximation, leading to artificial diffusion of the radiation, unphysical self-interaction, and other artificial effects. These methods typically include the time-dependent term of the RTE and sometimes artificially decrease the speed of light to reduce computational costs. Their computational expenses are independent of the number of sources, making them well-suited for calculating diffuse radiation. Frequency-wise, moment-based methods often use the grey approximation or, occasionally, the multi-group frequency approach. They are sometimes combined with ray tracing, which calculates radiation from one or several discrete sources, while the moment-based method addresses the diffuse radiation component. The most commonly used methods of this type include flux-limited diffusion, the M1 closure, and the variable Eddington tensor method.

Methods that directly solve the RTE integrate it along specific lines, commonly referred to as rays. These methods include forward and reverse ray tracing, long, hybrid, and short characteristics, as well as Monte Carlo methods. Ray tracing and characteristics-based methods typically do not include the time-dependent RTE term, utilizing the infinite speed of light approximation; however, there are exceptions. Monte Carlo methods are usually time-dependent. Depending on angular discretization, ray tracing and long/hybrid characteristics can achieve high accuracy but at the expense of high computational costs. Short characteristics methods tend to be more diffusive. While Monte Carlo methods can be highly accurate, achieving this requires a substantial number of photon packets, leading to increased computational costs. Monte Carlo methods are also versatile,  suitable e.g. for treating radiation scattering. Computational costs for most of these methods scale with the number of sources, with exceptions like reverse ray tracing methods that group distant sources and methods combining ray tracing with advection schemes. Various treatments of radiation frequency dependence are employed; generally, methods designed for a small number of sources can incorporate detailed frequency dependence. Forward ray tracing and long characteristics methods are typically challenging to parallelize using domain decomposition, whereas short characteristics are relatively straightforward. Hybrid characteristics, reverse ray tracing, and Monte Carlo methods fall in the middle ground regarding parallelization difficulty.

Radiation transport has become a well-established component of star formation simulations, with many methods and codes available, continuously advancing in both maturity and complexity. Consequently, it becomes imperative to not only validate these codes against standard tests but also to benchmark them against each other and, ideally, against realistic problems inherent in the realm of star formation. In this regard, initiatives such as the code comparison projects, such as those conducted by \citet{2006Iliev,2009Iliev} for cosmological radiation transport codes and the STARBENCH project led by \citet{2015BisbasB} that compares radiation transport codes for star formation, prove to be very valuable. Moreover, there is a growing need for more comparison projects to enhance our understanding and confidence in the capabilities of these codes.

\section*{Conflict of Interest Statement}

The authors declare that the research was conducted in the absence of any commercial or financial relationships that could be construed as a potential conflict of interest.



\section*{Funding}
This study has been supported by the institutional  project  RVO:67985815.

\section*{Acknowledgments}
I thank the anonymous referees for their constructive comments that have helped to improve this work. This study has been supported by the institutional  project  RVO:67985815.



\bibliographystyle{Frontiers-Harvard} 
\bibliography{refs}




\section*{Tables}

\end{document}